\pdfoutput=1
\documentclass[amsmath,prd,aps,twocolumn,showpacs,superscriptaddress,floatfix,nofootinbib]{revtex4}

\usepackage{amsfonts}
\usepackage{bm}
\usepackage{color}
\usepackage{graphicx}
\definecolor{darkgreen}{cmyk}{0.85,0.2,1.00,0.2}

\newcommand{\be}{\begin{equation}}
\newcommand{\ee}{\end{equation}}
\newcommand{\ba}{\begin{eqnarray}}
\newcommand{\ea}{\end{eqnarray}}
\newcommand{\nn}{\nonumber}

\newcommand\lsim{\mathrel{\rlap{\lower4pt\hbox{\hskip1pt$\sim$}}
        \raise1pt\hbox{$<$}}}
\newcommand\gsim{\mathrel{\rlap{\lower4pt\hbox{\hskip1pt$\sim$}}
        \raise1pt\hbox{$>$}}}

\def\Msol{M_\odot}
\def\r{{\bf r}}
\def\n{{\bf \widehat n}}
\def\k{{\bf k}}
\def\ellmax{\ell_{\rm max}}
\def\threej#1#2#3#4#5#6{\left( \begin{array}{ccc} #1 & #2 & #3 \\ #4 & #5 & #6 \end{array} \right)}

\begin{document}

\title{Supersonic baryon-CDM velocities and CMB B-mode polarization}

\author{Simone Ferraro}
\affiliation{Princeton University Observatory, Peyton Hall, Ivy Lane, Princeton, NJ 08544 USA}
\author{Kendrick M.~Smith}
\affiliation{Princeton University Observatory, Peyton Hall, Ivy Lane, Princeton, NJ 08544 USA}
\author{Cora Dvorkin}
\affiliation{Kavli Institute for Cosmological Physics, Enrico Fermi Institute, University of Chicago, Chicago, IL 60637}
\affiliation{Department of Physics, University of Chicago, Chicago, IL 60637}
\affiliation{School of Natural Sciences, Institute for Advanced Study, Princeton, NJ 08540}

\date{\today}

%%%%%%%%%%%%%%%%%%%%%%%%%%%%%%%%%%%%%%%%%%%%%%%%%%%%%%%

\begin{abstract}
It has recently been shown that supersonic relative velocities between dark matter
and baryonic matter can have a significant effect on formation of the first structures
in the universe.
If this effect is still non-negligible during the epoch of hydrogen reionization, it generates large-scale
anisotropy in the free electron density, which gives rise to a CMB B-mode.
We compute the B-mode power spectrum and find a characteristic shape with acoustic
peaks at $\ell\approx 200, 400, \ldots$.
The amplitude of this signal is a free parameter which is related to the dependence of the
ionization fraction on the relative baryon-CDM velocity during the epoch of reionization.
However, we find that the B-mode signal is undetectably small
for currently favored reionization models in which hydrogen is reionized promptly
at $z\sim 10$, although constraints on this signal by future experiments may help
constrain models in which partial reionization occurs at higher redshift, e.g.~by accretion
onto primordial black holes.
\end{abstract}

% \pacs{98.80.-k,~98.65.-r,~98.80.Cq,~95.36.+x}

\maketitle

%%%%%%%%%%%%%%%%%%%%%%%%%%%%%%%%%%%%%%%%%%%%%%%%%%%%%%%
%%%%%%%%%%%%%%%%%%%%%%%%%%%%%%%%%%%%%%%%%%%%%%%%%%%%%%%

\section{Introduction}

It was recently realized \cite{Tseliakhovich:2010bj}
that supersonic relative velocities between baryons and cold dark matter
can have a significant effect on formation of the first structures
in the universe.
The physics can be described intuitively as follows.

\begin{figure}
\centerline{\includegraphics[width=8cm]{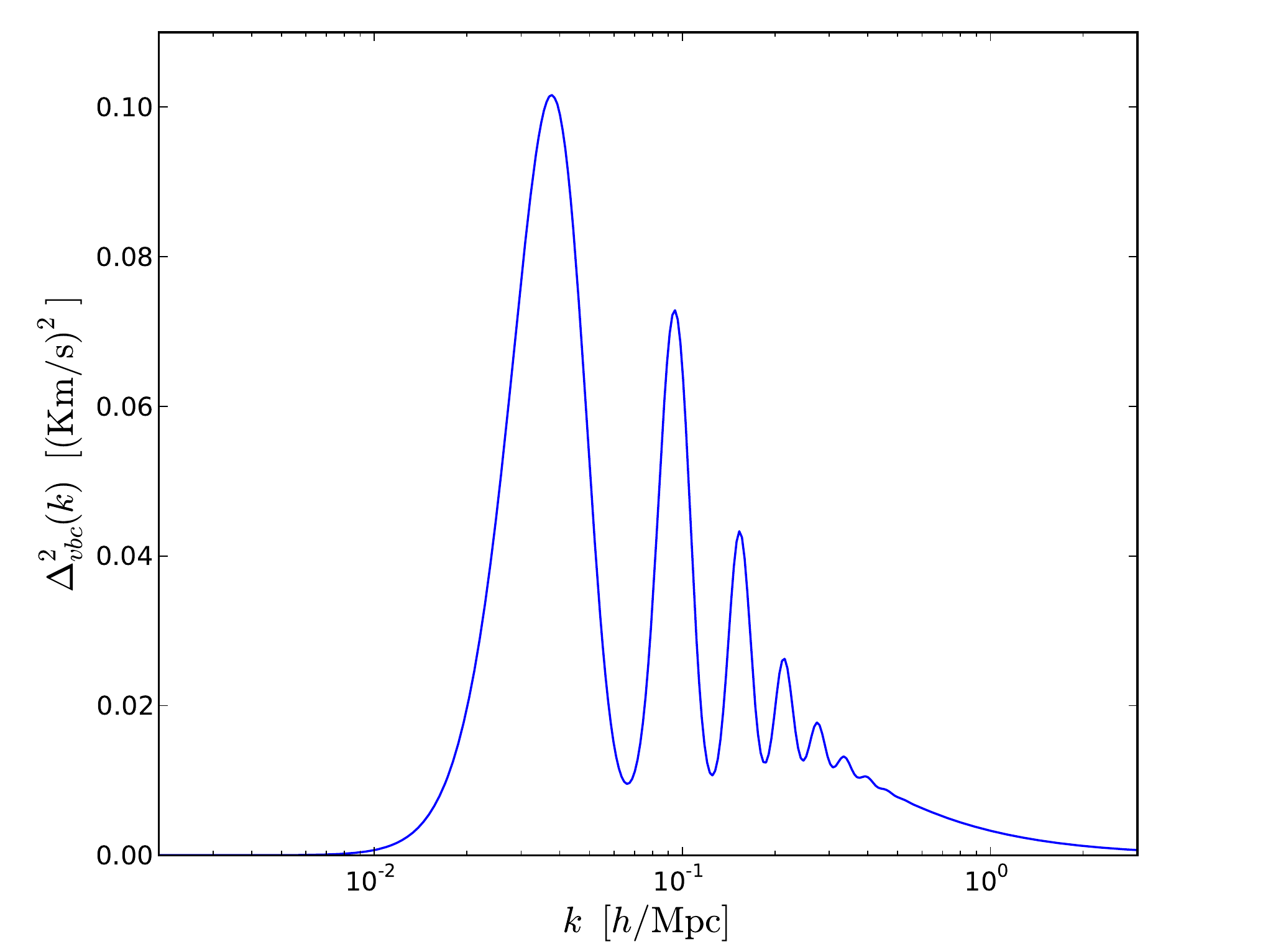}}
\caption{Power spectrum of the baryon-CDM relative velocity field $v_{bc}$ at $z=10.5$.
The power spectrum drops sharply for $k\gsim 1$ $h$ Mpc$^{-1}$, indicating
that the baryon-CDM flow is coherent on scales larger than a megaparsec.  The shape
of the power spectrum is roughly independent of $z$ (for $z\lsim 100$), but its 
amplitude varies as $\Delta^2_{vbc}(k)=k^3P_{vbc}(k)/2 \pi^2 \propto (1+z)^2$, with RMS velocity 
$\langle v_{bc}^2 \rangle^{1/2} = 0.36$ Km/s at $z=10.5$.}
\label{fig:relvel}
\end{figure}

The baryon-CDM relative velocity $v_{bc}({\bf x})$ is coherent on
scales smaller than a few comoving Mpc (Fig.~\ref{fig:relvel}).
Let us imagine dividing the universe into subregions of this size, 
with a constant value of $v_{bc}$ in each subregion.
As shown in \cite{Tseliakhovich:2010bj}, 
the baryon-CDM flow will suppress growth of perturbations whose comoving 
wavenumber $k$ is larger than the freestreaming scale $k_{FS} = aH/v_{bc}$.
Equivalently, a mode is suppressed if its wavelength is less than
the distance that the baryon-CDM flow travels in a Hubble time.

\begin{figure}
\centerline{\includegraphics[width=8cm]{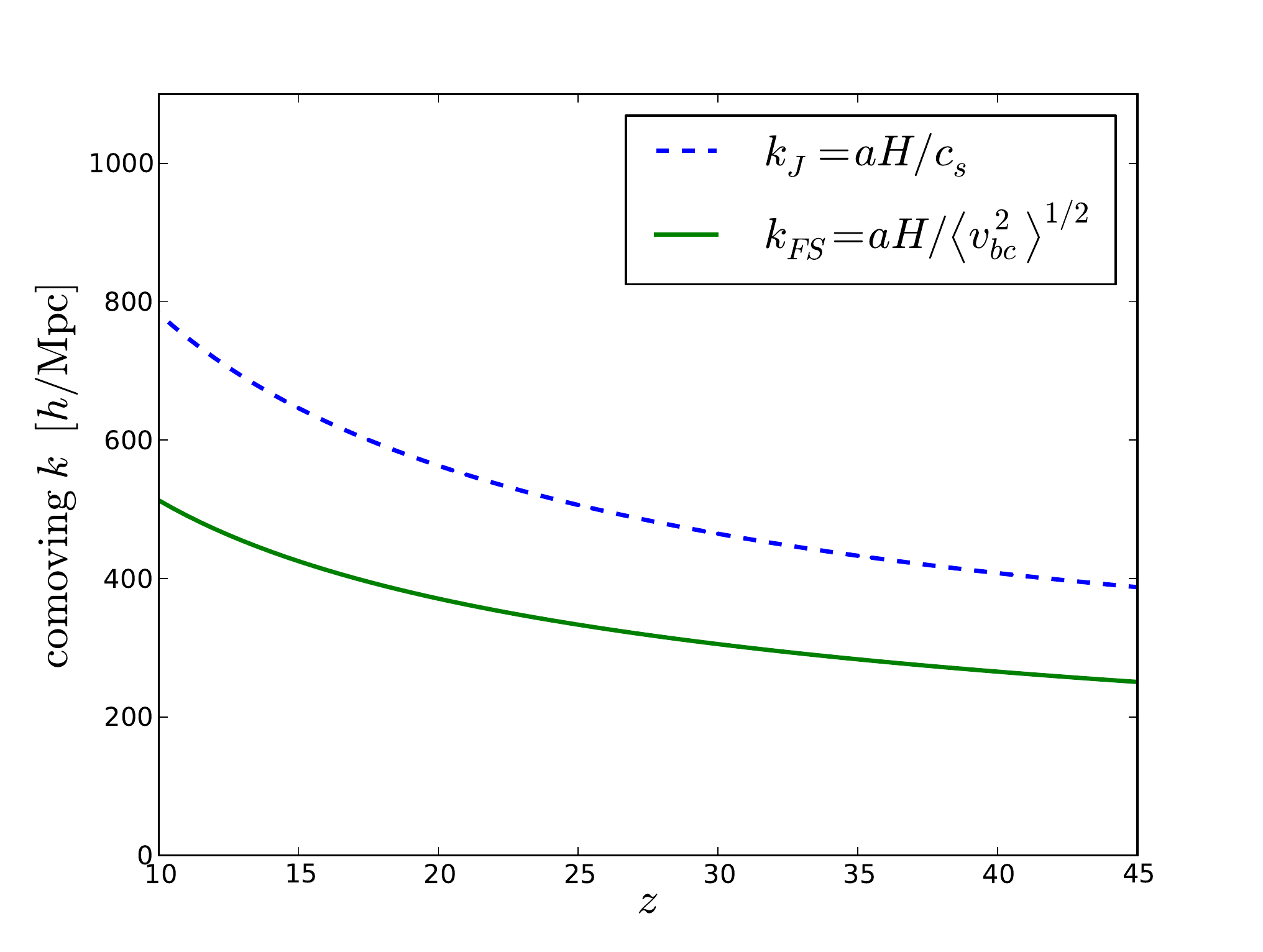}}
\caption{Redshift evolution of length scales relevant for the $v_{bc}$ effect.
As described in the text, the baryon-CDM flow suppresses structure formation
on a range of scales given by $k_{FS} \lsim k \lsim k_J$.
This makes the abundance of halos of mass $\lsim 10^6$ $\Msol$ sensitive
to the local value of $v_{bc}$, but the effect decreases quickly with increasing
halo mass.
}
\label{fig:comovinglength}
\end{figure}

Perturbations whose wavenumber $k$ is larger than the Jeans scale
$k_J = aH/c_s$, where $c_s$ is the baryon sound speed, are additionally
suppressed by the usual Jeans mechanism (pressure support).  For large $k$,
Jeans suppression dominates (by an extra factor of $k$ in the equations
of motion) over the $v_{bc}$ suppression.  Therefore, the $v_{bc}$ suppression is
important on a window of scales given by $k_{FS} \lsim k \lsim k_J$,
illustrated in Fig.~\ref{fig:comovinglength}.
The size $(k_J/k_{FS})$ of this window is equal to ${\mathcal M} = v_{bc}/c_s$, the Mach
number of the flow, so the existence of a range of scales where the $v_{bc}$
suppression dominates is equivalent to a supersonic flow.
For $z\lsim 100$, a typical Mach number is ${\mathcal M} = \langle v_{bc}^2 \rangle^{1/2} / c_s \approx 1.5$,
so there exists a range of scales where $v_{bc}$ suppression is an order unity effect.

The first stars in the universe are predicted to form at redshift $z\sim 30$, in
dark matter halos with typical mass $M\sim 10^6$ $\Msol$.
The abundance of such halos depends sensitively on the amplitude of perturbations
on a broad range of scales near characteristic wavenumber
$k = (4\pi\rho_m/3M)^{1/3} \sim 80h$ Mpc$^{-1}$.
This range of scales is broad enough that there is some overlap with the range 
$k_{FS} \lsim k \lsim k_J$ where $v_{bc}$
suppression is important, and so the abundance of the first stars is sensitive to the
local value of $v_{bc}$.
For larger halo masses, the overlap of scales decreases rapidly, and $v_{bc}$ suppression
is less important.

If we spatially average over a large region of the universe, then the overall effect
of the baryon-CDM flow is to delay the formation of the first bound structures.
What is more distinctive is that the effect is inhomogeneous: the
halo abundance will depend on the local value of $v_{bc}$.
Since the power spectrum of the $v_{bc}$ field has a different shape than the usual 
contributions to the halo power spectrum (i.e.~a Poisson term and a term proportional to the
matter power spectrum), this makes the $v_{bc}$ effect qualitatively new, e.g.~it
was shown in \cite{Yoo:2011tq} that the $v_{bc}$ effect can shift the apparent scale
of the baryon acoustic peak in the halo two-point function.

At redshift $z\sim 30$, the abundance
of halos (or stars) is significantly modulated by the local value of $v_{bc}$.
What is less clear (and a critical question for the phenomenology)
is whether the effect ``survives'' to lower redshift: if we
inspect the universe at $z=10$, can we still tell the difference between a low-$v_{bc}$
and a high-$v_{bc}$ region?
The gravitational suppression mechanism described above suggests that the abundance of
the largest halos is relatively unaffected by $v_{bc}$: the scales which are
suppressed by $v_{bc}$ are smaller than the scales which are relevant for spherical
collapse of these halos (Fig.~\ref{fig:comovinglength}).
However, the number 
(or total  luminosity) of luminous objects such as galaxies could depend on $v_{bc}$, due to
a variety of possible feedback effects.
An example of positive feedback would be metal enrichment of the intergalactic medium: 
in a large-$v_{bc}$ region, formation of the first stars is suppressed, leading to lower IGM metallicity,
which further suppresses star formation at later times since it will be harder
for gas to cool via atomic lines.
In the opposite direction, an example of negative feedback would be heating of the IGM
by starlight: in a large-$v_{bc}$ region with fewer stars at $z\sim 30$, the IGM will
be colder than average, which makes it easier
for clouds of gas to overcome pressure support and collapse gravitationally, leading
to increased star formation at later times.

It is also interesting to take an agnostic approach to the question of whether the $v_{bc}$
effect survives to low redshift (i.e.~treating the size of the effect as a free parameter),
and study the potential impact on various cosmological observables.
For example, a recent paper \cite{Yoo:2011tq} considered the possibility that the $v_{bc}$ effect survives
long enough to affect galaxy abundance at redshifts $z\lsim 5$.
Here, we will study the analogous possibility that the $v_{bc}$ effect survives long
enough to affect the hydrogen ionization fraction during the epoch of reionization $(z\sim 10)$,
and compute the CMB B-mode which is generated.
In particular, we would like to understand whether large-scale B-modes sourced by the $v_{bc}$ effect
can significantly contaminate the gravity wave signal at $\ell\approx 50$, since this will be a primary
scientific target for polarization experiments in the near future.

\section{B-mode calculation}

We assume that the mean ionization fraction $x_e(z)$ at redshift $z$ depends weakly on
the local value of $v_{bc}$.  By rotation invariance, $x_e$ can only depend on $v_{bc}^2$;
it is convenient to change variables by defining the field
\be
\eta(\r) = \frac{v_{bc}^2(\r,z)}{\langle v_{bc}^2(z) \rangle} - 1  \label{eq:eta_def}
\ee
which is normalized to have zero mean, and is (to an excellent approximation) independent of 
redshift as implied by the notation.

We will assume that the $v_{bc}$ effect is small at $z\sim 10$, so that
the ionization fraction $x_e(z)$ can be linearized in $\eta$:
\be
x_e(\r,z) = \bar x_e(z) \left( 1 + b_{xe}(z) \eta(\r) \right) \label{eq:xemodel}
\ee
where we have introduced a free parameter $b_{xe}(z)$ which parameterizes the size of
the $v_{bc}$ effect at redshift $z$.
Intuitively, $b_{xe}(z)$ is the mean fractional change in $x_e(z)$ produced by a $1\sigma$
fluctuation in the local value of $v_{bc}$.
The CMB optical depth in direction $\n$ is given by
\be
\tau(\n) = n_{e,0} \sigma_T \int dz\, \frac{(1+z)^2}{H(z)} x_e(\chi(z)\n, z)
\ee
where $n_{e,0}$ is the comoving electron density at redshift zero, $\sigma_T$ is the Thomson scattering cross section,
and $\chi(z)$ is the comoving distance to redshift $z$ in the assumed flat cosmology.
In the model~(\ref{eq:xemodel}), the angular power spectrum of the $\tau$ field is given (in the Limber approximation) by
\be
\frac{\ell^2 C_\ell^{\tau\tau}}{2\pi} = \int dz\, H(z) \left( \frac{d\bar\tau}{dz} \right)^2 b^2_{xe}(z) 
  \left( \frac{k^2 P_\eta(k)}{2\pi} \right)_{k=\ell/\chi(z)}  \label{eq:cltau_z_integral}
\ee
Note that we are only considering new contributions to $C_\ell^{\tau\tau}$ due to variations in the ionization
fraction due to the $v_{bc}$ effect; the full power spectrum contains additional terms (e.g.~a ``one-bubble'' term
arising from Poisson statistics of individual HII regions) not considered here \cite{Mortonson:2006re}.

Linear CMB polarization is generated by Thomson scattering of CMB photons by free electrons.
If reionization is spatially homogeneous, then the combination of rotation and parity invariance implies that
only E-mode polarization is generated.
The CMB B-mode power spectrum generated by inhomogeneous reionization is a sum of scattering and screening terms
\cite{Hu:1999vq,Dvorkin:2008tf,Dvorkin:2009ah}; these correspond respectively to generation of new linear polarization via Thomson
scattering by free electrons, and screening of the primary E-mode by inhomogeneities in the optical depth.
\ba
C_\ell^{BB} &=& C_\ell^{B_{\rm sca}} + C_\ell^{B_{\rm scr}} \\
C_\ell^{B_{\rm sca}} &=& \frac{3}{100} e^{-2\tau_{\rm eff}} Q_{\rm rms}^2 C_\ell^{\tau\tau}
\ea
\ba
C_\ell^{B_{\rm scr}} &=& e^{-2\tau_{\rm eff}} \sum_{\ell'\ell''} 
  \frac{1-(-1)^{\ell+\ell'+\ell''}}{2}
  \frac{(2\ell'+1)(2\ell''+1)}{4\pi} \nn \\
 && \hspace{0.3cm} \times
  \threej{\ell}{\ell'}{\ell''}{-2}{2}{0}^2
  C_{\ell'}^{EE(rec)} C_{\ell''}^{\tau\tau}
\ea
where $\tau_{\rm eff} \approx 0.09$ is the effective optical depth to reionization and $Q_{\rm rms} = 17.7$ $\mu$K is the RMS
temperature quadrupole during matter domination.

\begin{figure}
\centerline{\includegraphics[width=8cm]{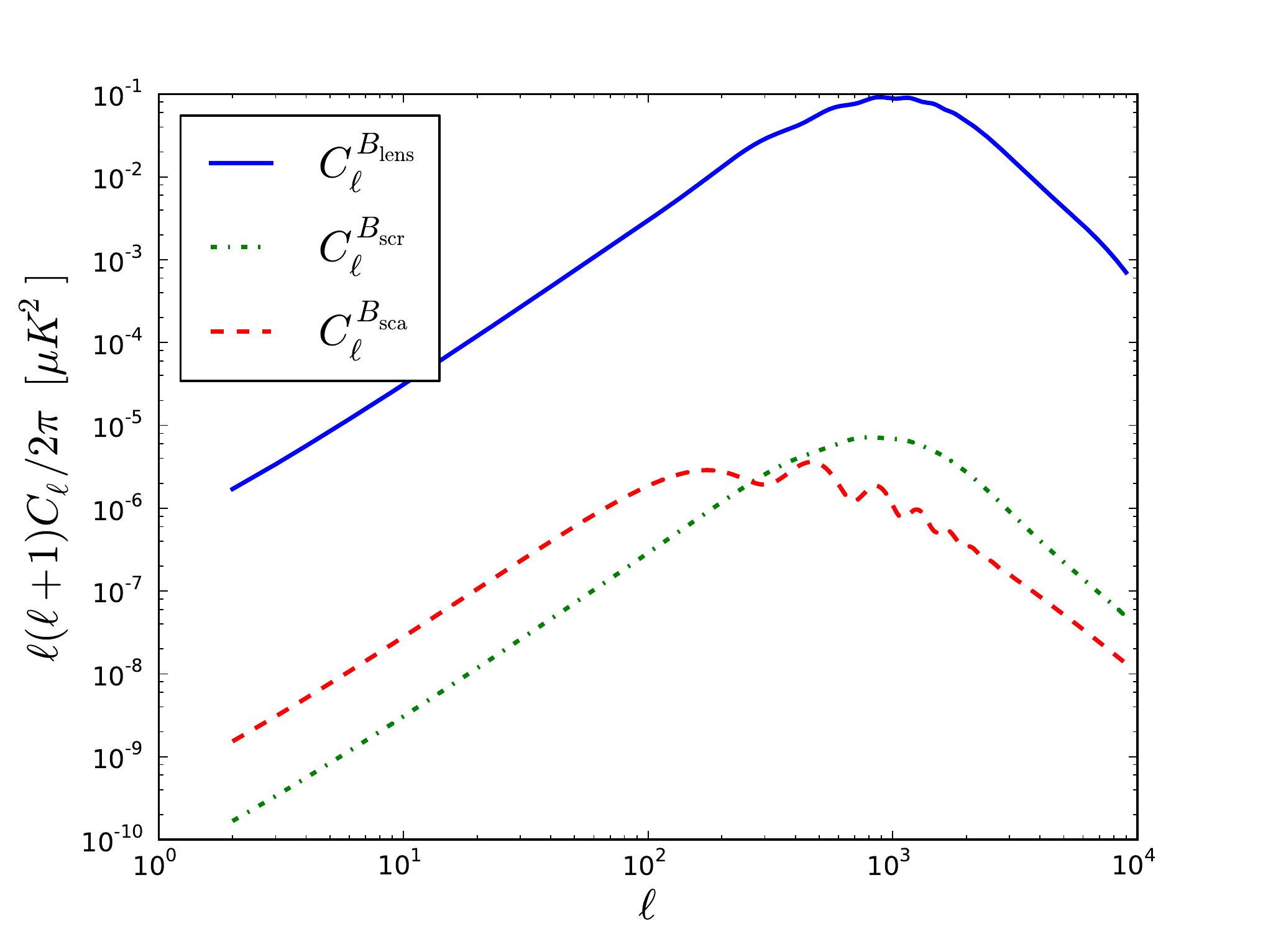}}
\caption{B-mode power spectra $C_\ell^{B_{\rm sca}}$ and $C_\ell^{B_{\rm scr}}$ generated by
baryon-CDM relative velocities, with the lensing B-mode power spectrum shown for comparison.
For $\ell \gsim 1000$, the B-mode power spectra $C_\ell^{B_{\rm sca}}, C_\ell^{B_{\rm scr}}$ will
be additionally suppressed (beyond what is shown here) by photon freestreaming during the
epoch of reionization.
For $\ell \lsim 1000$, these power spectra have shapes which are roughly independent
of the reionization model, but the overall amplitude $\alpha$ is a free parameter.
In the figure, we used $\alpha = 8.6 \times 10^{-5}$, corresponding to the fiducial reionization
model described in the text.}
\label{fig:clbb}
\end{figure}

For calculating the B-mode power spectrum at the $\approx$10\% level, we can make the approximation that the
comoving distance $\chi(z)$ is constant over the range of redshifts which contribute to the integral in~(\ref{eq:cltau_z_integral}).
We can then factor $C_\ell^{\tau\tau}$ as the product of a parameter $\alpha$ which depends on reionization history and the
bias parameter $b_{xe}(z)$, and a template shape in $\ell$ which depends only on cosmological parameters:
\be
C_\ell^{\tau\tau} \approx \alpha C_\ell^{\rm templ}
\ee
\be
\alpha = \int dz \frac{H(z)}{H_0} \left( \frac{d\bar\tau}{dz} \right)^2 b^2_{xe}(z)
\ee
\be
\frac{\ell^2 C_\ell^{\rm templ}}{2\pi} = H_0 \left( \frac{k^2 P_\eta(k)}{2\pi} \right)_{k=\ell/\chi_{\rm rei}} 
\ee
To complete the calculation of the B-mode power spectrum, we need to compute the power spectrum $P_\eta(k)$.
The two-point correlation function of the vector field $v_{bc}^i$ is given by:
\be
\langle v_{bc}^i(\k)^* v_{bc}^j(\k') \rangle = \frac{k^i k^j}{k^2} P_{vbc}(k) (2\pi)^3 \delta^3(\k-\k')   \label{eq:pk_vij}
\ee
where $i,j$ are spatial indices and $P_{vbc}(k)$ is the relative velocity power spectrum, which can be computed using CAMB \cite{Lewis:1999bs}.
From the definition~(\ref{eq:eta_def}) of $\eta$, the power spectrum $P_\eta(k)$ is a four-point function in $v_{bc}$
which can be computed straightforwardly using~(\ref{eq:pk_vij}) and Wick's theorem.
The result can be presented either in Fourier space,
\ba
P_\eta(k) &=& \frac{1}{4\pi^2 k \langle v_{bc}^2 \rangle^2} \int_0^\infty dk_1\, \int_{\max(k_1,|k-k_1|)}^{k+k_1} dk_2\, \nn \\
  && \hspace{0.5cm} \times \frac{(k^2 - k_1^2 - k_2^2)^2}{k_1 k_2} P_{vbc}(k_1) P_{vbc}(k_2)
\ea
or equivalently in position space as
\be
P_\eta(k) = 4\pi \int dr\, r^2 j_0(kr) \left( 6 \psi_0(r)^2 + 3\psi_2(r)^2 \right)
\ee
where the correlation functions $\psi_0(r)$, $\psi_2(r)$ are defined by:
\ba
\psi_0(r) &=& \frac{1}{3 \langle v_{bc}^2 \rangle} \int \frac{k^2\, dk}{2\pi^2} P_{vbc}(k) j_0(kr) \\
\psi_2(r) &=& -\frac{2}{3 \langle v_{bc}^2 \rangle} \int \frac{k^2\, dk}{2\pi^2} P_{vbc}(k) j_2(kr)
\ea
In Fig.~\ref{fig:clbb}, we show the B-mode power spectrum for a fiducial reionization model defined
as follows.
The reionization history $x_e(z)$ is given by the CAMB parameterization, with reionization width
$\Delta z=1.5$, central redshift $z_{\rm rei}=10.5$, and total optical depth $\tau=0.09$.
To get a sense for the order of magnitude of $\alpha$, if $b_{xe}(z)$ is constant in $z$, then 
$\alpha = 0.0086 b_{xe}^2$.
In \cite{Dalal:2010yt}, the value of $b_{xe}^2$ at $z\sim 10$ is predicted to be $\approx$ 0.01
(this value is consistent with the first attempts to include the $v_{bc}$ effect in simulations
\cite{Maio:2010qi,Stacy:2010gg,Greif:2011iv,Naoz:2011if}),
so we will take $\alpha = 8.6 \times 10^{-5}$ as our fiducial value.

For this fiducial reionization model, we find that the total B-mode power spectrum generated by baryon-CDM relative velocities is small
(roughly 0.14 $\mu$K-arcmin on large scales).
The scattering power spectrum $C_\ell^{B_{\rm sca}}$ has acoustic peaks at values of $\ell$ which are multiples of 
$\ell_* = \pi \chi_{\rm rei} / s_{\rm BAO}$, where $\chi_{\rm rei} \approx 9800$ Mpc is the angular diameter distance to
reionization and $s_{\rm BAO} \approx 150$ Mpc is the baryon acoustic scale.

There is one additional effect which has not been included in our B-mode calculation: high-$\ell$ damping due to freestreaming.
Our model~(\ref{eq:xemodel}) allows the mean ionization fraction $x_e$ at position $\r$ to depend on the velocity $v_{bc}(\r)$ at the same
point $\r$; in reality it will also depend on nearby values of $v_{bc}$ since the mean free path of reionizing photons is nonzero.
The qualitative effect will be to suppress the power spectra $C_\ell^{B_{\rm sca}}, C_\ell^{B_{\rm scr}}$ on scales
$\ell \gsim (\chi_{\rm rei} / \lambda) \approx 1000$, where $\lambda \approx 10$ Mpc is the 
characteristic size of HII regions during reionization, but a precise calculation of the damping tail
will depend on model-dependent details.

\section{Observational prospects and discussion}

To forecast detectability of the CMB B-modes produced by the $v_{bc}$ effect,
we consider several observational scenarios.
The $1\sigma$ error on $\alpha$ is given by
\be
\sigma(\alpha) = \left[ \frac{f_{\rm sky}}{2} \sum_\ell (2\ell+1) \left( 
  \frac{\partial C_\ell^{B_{\rm lens}}  / \partial \alpha}{C_{\ell}^{BB} + N_\ell^{BB}}
\right)^2 \right]^{-1/2}
\ee
where $C_\ell^{B_{\rm lens}}$ is the lensing B-mode power spectrum and
$N_\ell^{BB}$ is the noise power spectrum of the experiment.

For a lensing-limited experiment (i.e.~$N_\ell^{BB} \ll C_\ell^{B_{\rm lens}}$) with $f_{\rm sky}=0.7$
and $\ellmax=1000$, we find $\sigma(\alpha) = 8.6 \times 10^{-4}$.
This forecast effectively treats the lensing B-mode as an extra source of noise and could potentially be improved by
incorporating delensing techniques (e.g.~\cite{Knox:2002pe,Kesden:2002ku,Hirata:2003ka}) 
which statistically separate the lensing B-mode from other components.
Using the forecasting methodology from \cite{Smith:2010gu}, we find that the statistical error can be improved
to $\sigma(\alpha) = 7.5 \times 10^{-4}$, $2.7 \times 10^{-4}$, or $1.3 \times 10^{-4}$, assuming a Gaussian
beam with $\theta_{\rm FWHM} = 3'$ and polarization noise level $\Delta_P$ = 3, 1, or 0.25 $\mu$K-arcmin respectively.
In all cases, the statistical error $\sigma(\alpha)$ is larger than the value $\alpha_{\rm fid} = 8.6\times 10^{-5}$,
indicating that no statistically significant detection of the $v_{bc}$ signal is possible in our fiducial reionization model.

The basic reason that the $v_{bc}$ B-mode is small in our fiducial model is that we have assumed
that reionization is dominated by stars at $z\lsim 10$, and that the abundance of these sources
depends weakly on the local value of $v_{bc}$ (i.e.~$b_{xe}$ is small at these redshifts).
A larger B-mode may be generated if $b_{xe}$ can become large.
For example, if positive feedback mechanisms (such as metal enrichment of the IGM) help the
$v_{bc}$ effect survive to $z\sim 10$, then $b_{xe}$ may be enhanced, leading to a larger signal.
Models of reionization in which high redshifts contribute non-negligible optical depth should
also lead to increased $b_{xe}$.
For example, scenarios where X-ray emission from primordial black holes ionize the universe
by a few percent at redshift $\sim 100$ have been considered e.g.~in \cite{Ricotti:2007au}.
Although a quantitative study is beyond the scope of this paper, the bias parameter $b_{xe}$ could
be as large as order unity in these models, since the Hoyle-Bondi rate for accretion of gas onto a black
hole is proportional to $(v_{bc} + c_s)^{-3/2}$, and $v_{bc}$ and $c_s$ are of the same order of
magnitude (so that statistical fluctuations in $v_{bc}$ lead to order-unity variations in the
accretion rate).

In conclusion, although optical depth anisotropy sourced by baryon-CDM relative velocities
generates a CMB B-mode in principle, we find that the amplitude is likely to be unobservably
small, if currently favored reionization models are correct.
On the other hand, no new data analysis is required to fit for the amplitude $\alpha$ of the 
$v_{bc}$ B-mode (one simply fits a multiple of a fixed template shape to the estimated power spectrum), 
so estimating $\alpha$ from future B-mode measurements may be a useful test of the assumptions
underlying these models.
In any case, since the acoustic features at $\ell=200,400,\ldots$ are on different scales than
the ones which are relevant for gravity waves ($\ell\lsim 10$ and $\ell\approx 50$), the B-mode
power spectrum calculated here cannot be a contaminant for the gravity wave signal, even in a reionization model 
where its amplitude is large.

%%%%%%%%%%%%%%%%%%%%%%%%%%%%%%%%%%%%%%%%%%%%%%%%%%%%%%%%%%%%%%%%%%%%%%%%%%%%%%%%%%%%%%%%%%%%%%%%%%%%

\vspace{1cm}

{\em Acknowledgements.}
We thank Renyue Cen, Dan Grin, Jerry Ostriker, and David Spergel for useful discussions.
SF was supported by the Martin Schwarzschild Fund in Astronomy at Princeton University.
KMS was supported by a Lyman Spitzer fellowship in the Department of Astrophysical Sciences at Princeton University.
CD was supported by the Kavli Institute for Cosmological Physics (KICP) at the University of Chicago through grants NSF PHY-0114422 and NSF PHY-0551142 and an
 endowment from the Kavli Foundation and its founder Fred Kavli. CD is additionally supported by the Institute for Advanced Study through the NSF grant AST-0807444 and the Raymond and Beverly Sackler Funds.

\bibliographystyle{prsty}
\bibliography{taupeak_notes}

\end{document}